\documentclass[ prl,showpacs,amsfonts,amssymb,amsmath,twocolumn]{revtex4}
\usepackage{times}
\usepackage{graphicx}

\newsavebox{\GSIM}
\sbox{\GSIM}{{$\ \stackrel{\textstyle>}{\sim}\ $}}

\begin{document}

\title{Probing temperature and damping rates in Bose-Einstein condensates using ultraslow light experiments}

\author{D. C. Roberts} \affiliation{Laboratoire de Physique Statistique de l'Ecole normale sup\'erieure, Paris, France}

\begin{abstract}
We propose a method to probe Landau and Beliaev processes in dilute trapped atomic condensates with a multiple state structure using ultraslow light experimental configurations.  Under certain conditions, damping rates from these collisional processes are directly proportional to the dephasing rates, making it possible to determine damping rates through measurement of the dephasing.  In the ultraslow light systems we consider, Landau decay rates are enhanced at low momenta, which allows one to distinguish between Landau-dominated and Beliaev-dominated regimes at the same temperature.  Furthermore, the enhancement of Landau rates potentially provides a way to measure low temperatures ($T \ll T_c$) in dilute condensates more accurately than current methods permit.

\end{abstract}
 
\maketitle

Agreement between mean field approximations and the observed behavior of collective excitations in dilute Bose-Einstein condensates  has been widely verified in low temperature experiments \cite{expT0}.  However, mean field approximations cannot describe many features of higher temperature regimes in which damping effects of the excitations become significant.  Studying the damping rates of excitations in dilute Bose-Einstein condensates allows one to explore thermal field theories beyond the confines of mean field approximation.  

It has been shown that the damping rates of the collective excitations exhibit a strong temperature dependence  \cite{expT}, which has been attributed to Landau processes \cite{liu} - the collision of two excitations to produce an excitation of higher energy.  Experimental evidence also exists of the reverse occurring, whereby an excitation decays into two lower energy excitations through various processes - collectively known as Beliaev processes - such as the coupling of scissor modes \cite{foot}, the decay of Kelvin modes \cite{dalibard}, and the decay process signaled by the suppression of quasiparticle collisions \cite{katz}.  There has been significant experimental and theoretical interest in these damping rates (see for example \cite{Giorgini, liu2, Fedichev, stringari} and references therein), yet a fully consistent understanding of these effects remains to be achieved.  

In this letter, we propose a method of studying the Landau and Beliaev damping rates in dilute trapped atomic condensates with a multiple state structure using ultraslow light experimental configurations.  Landau processes occurring in the multiple-state internal atomic structure of these systems are enhanced at low excitation momenta compared to Landau processes in systems of atoms with a single internal state.  This enhancement makes apparent the relative contribution of the Landau and Beliaev processes to the overall decay rate as a function of excitation momenta, allowing three regimes to be identified:  1) dominance of Landau decay, 2) dominance of Beliaev decay, and 3) comparable contribution from both Landau and Beliaev decay processes.  The enhancement of Landau dissipation also provides a more effective way to measure very low temperatures ($T \ll T_c$, where $T_c$ is the critical temperature) for which, due to the lack of thermal atoms, the usual measure of the velocity distribution of an expanded condensate is not reliable.  

We consider a dilute homogeneous gas of bosons of mass $m$ confined in a box of volume V (the effect of the trapping potential on the homogeneity approximation will be discussed briefly below).   In the ultraslow light systems we consider, three relevant internal states exist: $|a\rangle$, $|b\rangle$ and $|c \rangle$.  Where it is pertinent to the discussion, we use
subscript notation to denote the  momentum of an excitation
within the associated internal state.
\begin{figure}[]
\begin{center}
\includegraphics[scale=0.7]{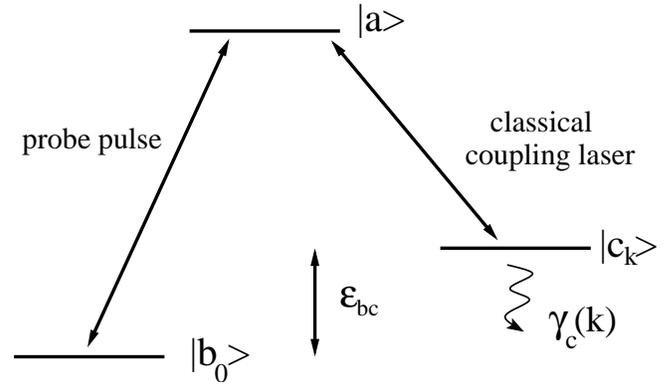}
\end{center}
\caption{\label{ramandecay} We consider 3-level atoms illuminated by a probe pulse  and a classical coupling laser.  We assume the condensate atoms to be in $|b_0\rangle$. The lasers excite the atoms, into the translational state $|c_k\rangle$.  $\gamma_c(k)$ represents the decay rate from $|c_k \rangle$.} 
\end{figure}
As seen in Figure \ref{ramandecay}, the ground state $|b_0 \rangle$, which is a weakly interacting Bose-Einstein condensate with $N_0$ atoms in the condensate, is resonantly coupled to  excited state $|a \rangle$ by a weak probe pulse; and the metastable translational state $|c_k \rangle$ of atoms with momentum $k$ is an energy $\epsilon_{bc}$ above $|b_0 \rangle$  and is resonantly coupled to $|a \rangle$ by a classical laser field.  Whereas in traditional EIT experiments the dark state is a superposition of the two lower atomic states and the light fields act as classical sources, in the present set-up the dark state is a superposition of the photons in the probe pulse and the atoms in $|c_k \rangle$  \cite{Fleishhauer2002a,roberts}, and it is the coupling pulse and the condensate that act as classical sources.  The present configuration has been shown experimentally to permit a dramatic slowing of light propagation to speeds on the order of 10 m/s \cite{Hau1999a} and the storage of light pulses for up to $1\,$ms before they are regenerated \cite{Liu2001a}. (For a recent review see \cite{Lukin2001a}.)  

If one assumes that 1) the pulse duration is sufficiently long that its bandwidth fits within the finite frequency width of the transparency window and 2) any incidental couplings to states outside this system can be ignored, the dephasing rates, which can be measured by the decay of the transmitted probe pulse energy, will arise primarily from atomic collisions \cite{Lukin2001a,Fleishhauer2002a}.  One can show that under typical experimental conditions the dephasing rates are proportional to $\gamma_c(k)$ \cite{roberts,janne}- the decay of the population of the state $|c_k \rangle$-  by a proportionality factor close to 1.0 (slightly greater for light delay experiments than for light storage experiments) \cite{roberts}.  

The population of state $|c_k \rangle$ decays mainly due to 2-body elastic spin-conserving collisions with the atoms in internal state $|b\rangle$ characterized by a positive scattering length $a_{bc}$.  To leading order, collisions between atoms in $|b\rangle$, characterized by a positive scattering length $a_{bb}$, change the free particle dispersion relation  to the Bogoliubov quasiparticle dispersion relation, $\epsilon^{b}_p=\mu p \sqrt{2+p^2}$ \cite{bog}, where $\mu=(\hbar k_0)^2/2m$ is the chemical potential and $p$ is the normalized excitation momentum.  (Unless otherwise specified all momenta are normalized by the healing length $k_0^{-1}=1/\sqrt{8 \pi n_0 a_{bc}}$ where $n_0=N_0/V$.) Higher order collisions among atoms in $|b\rangle$, which cause the quasiparticle populations to decay, can be ignored in the dark state picture because the atoms in $|b\rangle$ serve only as a classical source in our system.  To allow us to treat the atoms in $|b\rangle$ as if they are in thermal equilibrium, we assume that atomic collisions within internal state $|b\rangle$ occur at much shorter time scales than atomic collisions between $|b\rangle$ and $|c_k \rangle$, i.e. $a_{bb} \gg a_{bc}$ \cite{feshbach}.   Collisions among atoms in internal state $|c\rangle$, characterized by a positive scattering length $a_{cc}$, can also be ignored as there is no macroscopically occupied mode in $|c\rangle$. Consequently the dispersion relation for excitations in $|c\rangle$ is the free particle dispersion relation given by $\epsilon^c_p=\mu p^2$ (note that the total energy of excitations in $|c\rangle$ is $\epsilon^c_p+\epsilon_{bc}$) \cite{nc}. We also assume that $a_{bb}a_{cc}> a_{bc}^2$ in order to avoid phase separation \cite{colson}.  Finally, collisions involving atoms in $|a\rangle$ can be ignored because the population of $|a\rangle$ is negligible.   

Assuming the system of atoms is near thermal equilibrium, one can determine $\gamma_c(k)$ by calculating the imaginary part of the perturbed energy \cite{robertsthesis, Morgan1999a}, by using the Golden Rule \cite{stringari} or by taking a time-dependent approach \cite{ Gasenzer2002a, Giorgini}.  The decay rate of $|c_k \rangle$ comprises the Beliaev and Landau decay rates, i.e. $\gamma_c(k)=\gamma^{\mathrm B}_c(k)+\gamma^{\mathrm L}_c(k)$.  The Beliaev decay rate, corresponding to the collision between a particle in $|c_k \rangle$ and a condensate atom in $|b_0\rangle$ to produce one quasiparticle in $|b_j \rangle$ of energy $\epsilon^{b}_j$ and one particle excitation in $|c_i\rangle$ of energy $\epsilon^{c}_i$ such that $\epsilon_k^c=\epsilon_i^c+\epsilon_j^b$, is given by
\begin{equation}
\gamma^{\mathrm B}_c(k)=\frac{U_{bc}^2 N_0 \pi }{V^2} \sum_j (u_j-v_j)^2 \left[ (1+n^c_i+n^b_{j}) \delta(\epsilon^c_k-\epsilon^b_j-\epsilon^c_i) \right]
\end{equation}
where the interaction strength for the contact potential between atoms in $|b\rangle$ and $|c\rangle$ is given by $U_{bc}=4 \pi \hbar^2 a_{bc}/m$  where  $n_{j}^b=\frac{1}{\exp[\epsilon^b_{j}/k_B T]-1}$ and $n_{i}^c=\frac{1}{\exp[(\epsilon_{k}^c+\epsilon_{bc}-\epsilon^b_j)/k_{B}T]-1}$ are the thermal populations associated with $|b_j \rangle$ and $|c_i\rangle$, respectively.
The Landau decay rate is caused by the reverse process - a particle in $|c_k \rangle$ collides with a quasiparticle in $|b_j \rangle$ to produce a particle excitation in $|c_i\rangle$ with higher energy $\epsilon_i^c=\epsilon_k^c+\epsilon^b_j$ and a condensate atom in $|b_0 \rangle$ - and is given by
\begin{equation}
\gamma^{\mathrm L}_c(k)=\frac{U_{bc}^2 N_0 \pi }{V^2} \sum_j (u_j-v_j)^2 (n^b_j-n^c_i) \delta(\epsilon^c_k+\epsilon^b_j-\epsilon^c_i)
\end{equation}
where $
n_{i}^c=\frac{1}{\exp[(\epsilon_{k}^c+\epsilon_{bc}+\epsilon^b_j)/k_B T]-1}$.  (Note that the difference in $n_{i}^c$ between the two cases will become irrelevant, as discussed below.)  $u_j$ and $v_j$ are the quasiparticle amplitudes that characterize the Bogoliubov transformation \cite{bog} and the square of their difference is
given by $(u_p-v_p)^2=p^2/z_p^b$ where the  dimensionless energy scaled by the chemical potential is given by $z_{p}^b=p\sqrt{2+p^2}$.

As we are considering a homogeneous system, the sums can be converted into integrals to arrive, after some manipulation, at
\begin{equation}
\label{beliaev}
  \gamma^{\mathrm B}_c(k)=\sqrt{n_0 a_{bc}^3} \frac{\mu}{\hbar}\frac{\sqrt{8 \pi}}{k}\int_0^{k^2} dz^b_j 
      \left(1-\frac{1}{\sqrt{1+(z^b_j)^2}} \right) (1+n_j^b+n_i^c),
\end{equation}
and
\begin{equation}
\label{landau}
  \gamma^{\mathrm L}_c(k) =\sqrt{n_0 a_{bc}^3} \frac{\mu}{\hbar}\frac{\sqrt{8 \pi}}{k}\int_0^\infty dz^b_j 
      \left(1-\frac{1}{\sqrt{1+(z^b_j)^2}} \right) (n_j^b-n_i^c).
\end{equation}

Since $\epsilon_{bc}/k_B T \gg 1$ in typical experimental situations involving condensates (note that we assume $\epsilon_{bc}/k_B T \gg 1$ in this paper unless otherwise specified), the number of thermally excited atoms in state $|c\rangle$ is negligible, i.e. $n_i^c \approx 0$, implying
\begin{equation}
 \gamma^{\mathrm L}_c(k)= \sqrt{8 \pi}\sqrt{n_0 a_{bc}^3} \frac{\mu}{\hbar}\frac{f(\tau)}{k} 
\end{equation}
where the dimensionless temperature is given by $\tau=k_B T/\mu$ and $f(\tau)$ is a function only of temperature and is given by $\int_0^\infty dz^b_j \left(1-\frac{1}{\sqrt{1+(z^b_j)^2}} \right) n_j^b(\tau)$.
 This assumption of a large energy gap between states $|b\rangle$ and $|c\rangle$ implies  a large difference in the thermal occupation numbers in $|b\rangle$ and $|c\rangle$, which has a significant effect on the Landau decay rate of the low momentum excitations in $|c\rangle$, which now diverges as $1/k$ as $k \rightarrow 0$ (the assumption that the excitations have well-defined momenta breaks down at small momenta, an effect that we describe below). Compare this to the single state case where $\gamma^{\mathrm L}(k) \rightarrow 0$ as $k \rightarrow 0$, e.g. $\gamma^{\mathrm L}(k) \propto  k \tau^4 $ for $k\ll \tau \ll 1$ and $\gamma^{\mathrm L}(k) \propto k \tau$ for 
$\tau \gg 1$ and $k \ll 1$ (see \cite{morganthesis} and references therein).  At high momenta where $k \gg 1$, the Landau decay rate falls off as $1/k$ as in the single state case \cite{highk}.

\begin{figure}[]
\begin{center}
\includegraphics[scale=0.8]{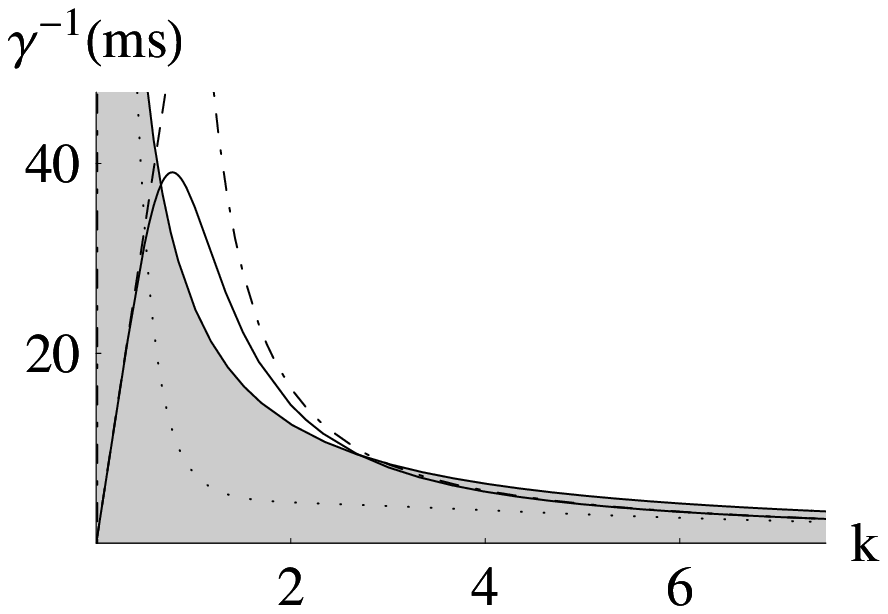}
\hfill
\includegraphics[scale=0.8]{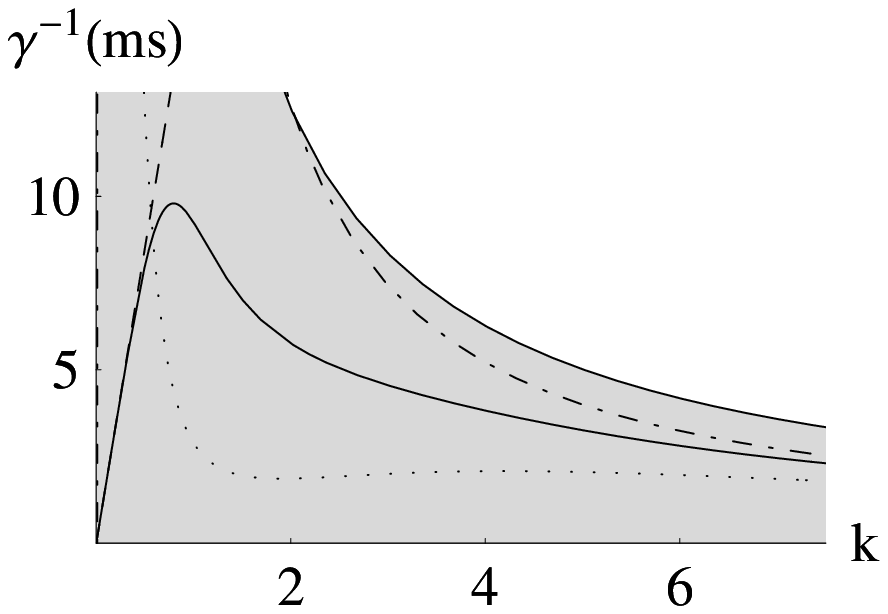}
\end{center}
\caption{\label{decay2} The inverse of the decay rate as a function of excitation momentum $k$ (scaled with the healing length) for $T=0.1 \hspace{0.5 mm} T_c$ ($\tau=0.86$) for the upper plot and $T=0.2 \hspace{0.5 mm} T_c$ ($\tau=1.72$) for the lower plot.  The solid line represents $1/\gamma_c(k)$ - the inverse of the total decay from $|c_k\rangle$ assuming the experimental realistic scenario of $\epsilon_{bc}/k_B T \gg 1$.  The dashed line represents the inverse of the decay due to Landau processes, $1/\gamma^{\mathrm L}_c(k)$, and the dot-dashed line represents the inverse of the decay due to Beliaev processes, $1/\gamma^{\mathrm B}_c(k)$, both assuming $\epsilon_{bc}/k_B T \gg 1$.  To illustrate the importance of $\epsilon_{bc}$ we include $1/\gamma_c(k)$ as a dotted line assuming $\epsilon_{bc}=0$.  The gray area signifies the excitation momenta that can be measured before the edge of the trap is reached, i.e. $1/\gamma_c(k) < \frac{\hbar k_0 R}{2 k \mu}$.  We assume typical experimental parameters: $a=3$ nm, $T_c=500$ nK, $\mu/\hbar = 2 \pi$ kHz, $k_0 = 1$ $\mu$m$^{-1}$, $R=100$ $\mu$m (along the major axis of elongated condensates).}
\end{figure}

The enhancement of Landau decay rates at low momenta in this system allows one to determine whether the Landau or Beliaev decay is dominant: the Landau decay rate decreases monotonically with momentum (unlike in the single state case) whereas the Beliaev decay rate increases monotonically with momentum as seen in Figure \ref{decay2}.  This allows one the possibility to independently probe Beliaev- and Landau-dominated processes at the same temperature.  Eventually, as the temperature increases, the Landau processes will become dominant over a wider range of momenta because Landau processes can involve any mode with an appreciable population whereas Beliaev processes are restricted to modes whose energies are less than that of decaying particles.  

We have assumed a homogeneous gas, which should be a good approximation for the central region of the large alkali-metal condensates achieved in present-day experiments.  However, we have implicitly assumed that excited atoms in state $|c_k\rangle$ remain in the interaction region, which is ultimately limited by the size of the condensate (for large condensates this is the Thomas-Fermi radius), for times longer than that of the characteristic decay, i.e. $\gamma_c(k) > 2 k \frac{\mu}{\hbar} \frac{1}{k_0 R}$ where $R$ is the characteristic size of the condensate.   This puts an upper limit on the excitation momentum one can measure; the measurable momenta are represented by the gray region in Figure \ref{decay2}.

We have implicitly assumed that the excitations have well-defined momenta, i.e. $\gamma_c(k)/\epsilon^c_k \ll 1$.  For the Landau-dominated regimes at low momenta this condition implies that $k^3 \gg \sqrt{8 \pi} \sqrt{n_0 a_{bc}^3} f(\tau)$, which  is not very restrictive because we assume a dilute gas such that $\sqrt{n_0 a_{bc}^3} \ll 1$, and $f(\tau) < 1$ for  temperatures in the range of $\tau <1$ where both Beliaev and Landau processes are relevant.   For the Beliaev-dominated regime at relatively high momenta ($k \gtrsim 1$) and  temperatures in the range of $\tau<1$, the zero-temperature result  $\gamma^{\mathrm B}_c(k)_{T=0}= g(k) \sqrt{n_0 a_{bc}^3}\mu/\hbar $ dominates; here, $g(k)=\sqrt{8 \pi} k [1-\ln(k^2+\sqrt{1+k^4})/k^2]$.    The assumption of well-defined momenta is satisfied when $\sqrt{n_0 a_{bc}^3} g(k)/k^2 \ll 1$, which is always true for a dilute gas since $g(k)/k^2 \lesssim 1$ for all $k$.

The enhancement of the Landau decay rate at low momenta has another important consequence: it offers a method of accurately determining very low temperatures, i.e. $T \ll T_c$.  At $T=0$ the Beliaev damping is the only remaining dissipative process, as is the case with the single-level system, and $\gamma_c(k)_{T=0}=   \gamma^{\mathrm B}_c(k)_{T=0} \rightarrow 0$ as $k \rightarrow 0$ as $k^5$ (the full expression is given above).    Unlike in the single atomic state case, however, at low temperatures $\tau= k_B T /\mu \ll 1$ Landau decay dominates at low momenta and can be written as
\begin{equation}
\gamma_c(k) \approx \gamma^{\mathrm L}_c(k) \approx \frac{\zeta(3)\sqrt{8 \pi}}{6} \sqrt{n_0 a_{bc}^3} \frac{\mu}{\hbar}   \frac{\tau^3}{k} 
\end{equation}
where $\zeta$ is the Riemann zeta function and the excitation momenta remain well defined for $k^3 \gg \sqrt{n_0 a_{bc}^3} \tau^3 $.

The decay approximation used in this paper is only strictly valid for our homogeneous system with its continuous spectrum of excitations (and perhaps also for anisotropic parabolic traps where the discrete energy levels are chaotically distributed \cite{Fedichev}). In experimental situations using a trap, however, the spectrum of excitations becomes significantly discretized at lower momenta, which would considerably impede Beliaev processes at low momenta.    
Nevertheless the argument remains unaffected that Landau decay rates are enhanced at low momenta due to the large energy gap between $|b \rangle$ and $|c \rangle$.  We maintain in this paper that this 1) makes it possible to distinguish between the Beliaev and Landau decay rates at the same temperature and 2) provides a sensitive low temperature thermometer.
   
The author gratefully acknowledges stimulating discussions with David Guery-Od\'elin and Sam Morgan.


\begin{thebibliography}{9}
\bibitem{feshbach} The different scattering lengths can be, in principle, tuned by Feshbach resonances \cite{fesh1}.
\bibitem{nc} We are assuming that the number of atoms in internal state $|c\rangle$ is less than  $\sqrt{N_0}$.
\bibitem{highk}  One can intuitively understand the disappearance of the Landau processes in the classical limit (ignoring bosonic stimulation) as follows \cite{morgan}: Beliaev processes can be thought of as the collision of a moving particle with a particle initially at rest (analogous to the condensate particle), resulting in the scattering of the colliding particles such that their onward trajectories are perpendicular to each other; Landau processes, being the reverse, require that the two classical particles be moving towards each other along perpendicular trajectories in order for the collision of the two particles to leave one particle at rest.  It is clear that the latter scenario occurs far more rarely than the former.  Indeed, at high momenta such that $k\gg1$ the effect of Beliaev processes at $T=0$ replicates the classically predicted rate of decay, i.e. $\gamma^{\mathrm B}_c(k) =\gamma_c(k)_{T=0} \approx \gamma_{KT}(k)$, where  $\gamma_{KT}(k)=n_0 \sigma v$ is the collision rate from classical kinetic theory (assuming the particles have speed $v=\hbar k_0 k/m$ and cross-sectional area $\sigma = 4 \pi a_{bc}^2$).
\end{thebibliography}
\end{document}